\documentclass[journal,10pt]{IEEEtran}
\usepackage{graphicx} 

\usepackage{cite}
\usepackage{amsmath,amssymb,amsfonts}
\usepackage{algorithmic,algorithmic}
\usepackage{graphicx}
\usepackage{textcomp}
\usepackage{xcolor}
\def\BibTeX{{\rm B\kern-.05em{\sc i\kern-.025em b}\kern-.08em
    T\kern-.1667em\lower.7ex\hbox{E}\kern-.125emX}}
\usepackage{graphicx}
\usepackage{subcaption}
\graphicspath{ {images/} }
\usepackage[utf8]{inputenc}
\usepackage{fancyhdr}
\usepackage{dirtytalk}
\usepackage{tabularx}
\PassOptionsToPackage{hyphens}{url}\usepackage{hyperref}
\usepackage[ruled,vlined]{algorithm2e}

\usepackage{enumitem,kantlipsum}
\usepackage{listings}
\usepackage{graphicx}
\graphicspath{ {images/} }
\usepackage[utf8]{inputenc}
\usepackage{fancyhdr}
\usepackage{dirtytalk}
\usepackage{tikz}
\usepackage{bbm}
\usepackage[most]{tcolorbox}

\usetikzlibrary{shapes.geometric, arrows}
\usepackage{pgfplots}
\pgfplotsset{width=8cm,compat=1.16}
\usepgfplotslibrary{external}
\usepackage{array}
\newcolumntype{P}[1]{>{\centering\arraybackslash}p{#1}}

\tikzstyle{startstop} = [rectangle, rounded corners, minimum width=2cm, minimum height=0.5cm,text centered, draw=black, fill=red!30]
\tikzstyle{process} = [rectangle, minimum width=2cm, minimum height=0.5cm, text centered, draw=black, fill=orange!30, align=left]
\tikzstyle{decision} = [diamond, minimum width=1.0cm, minimum height=0.4cm, text centered, draw=black, fill=green!30]
\tikzstyle{arrow} = [thick,->,>=stealth]

\newcolumntype{L}{>{$}l<{$}}

\usepackage[labelformat=simple]{subcaption}

\usepackage[nolist]{acronym}
\begin{acronym}
\acro{LM}{language model}
\acro{LLM}{large language model}
\acro{SLM}{small language model}
\acro{RAG}{retrieval augmented generation}
\acro{LM}{language model}
\acro{LoRA}{low-rank adaptation}
\acro{QLoRA}{quantized low-rank adaptation}
\acro{SBERT}{sentence BERT}
\acro{ML}{machine learning}
\acro{TF-IDF}{term frequency and inverse document frequency}
\acro{OOD}{out-of-distribution}
\acro{QnA}{questions and answering}
\acro{AI}{artificial intelligence}

\acro{3GPP}{third generation partnership project}
\acro{O-RAN}{open radio access networks}
\acro{MCQ}{multiple choice question}

\acro{MAS}{multi-agent system}
\acro{CRN}{cognitive radio network}

\acro{MDP}{Markov decision process}
\acro{IC}{incentive compatibility}
\acro{IR}{individual rationality}
\acro{URLLC}{ultra reliable low latency}
\acro{AI}{artificial intelligence}
\acro{SNR}{signal-to-noise-ratio}
\acro{CI/CD}{continuous integration and continuous deployment}
\acro{GAN}{generative adversarial network}
\acro{NFV}{network function virtualization}
\acro{SDN}{software-defined networking}
\acro{RLHF}{Reinforcement learning from human feedback}
\acro{QoS}{quality-of-service}
\acro{SINR}{signal-to-interference-plus-noise ratio}
\acro{RTB}{real time bidding}

\acro{VCG}{Vickrey–Clarke–Groves}
\acro{GFP}{Generalized First Price}

\acro{HetNets}{heterogeneous networks}
\acro{UE}{user equipment}
\acro{BS}{base station}

\end{acronym}

\begin{document}
\captionsetup[figure]{labelfont={bf},labelformat={default},labelsep=period,name={Figure}}

\title{Strategic Bidding in 6G Spectrum Auctions with Large Language Models}

\author{Ismail Lotfi and Ali Ghrayeb

\thanks{
The authors are with the College of Science and Engineering, Hamad Bin Khalifa University, Doha, Qatar. (e-mails: \{ilotfi,
aghrayeb\}@hbku.edu.qa).
}
}

\maketitle

\begin{abstract}
Efficient and fair spectrum allocation is a central challenge in 6G networks, where massive connectivity and heterogeneous services continuously compete for limited radio resources.
We investigate the use of Large Language Models (LLMs) as bidding agents in repeated 6G spectrum auctions with budget constraints in vehicular networks. Each user equipment (UE) acts as a rational player optimizing its long-term utility through repeated interactions. Using the Vickrey–Clarke–Groves (VCG) mechanism as a benchmark for incentive-compatible, dominant-strategy truthfulness, we compare LLM-guided bidding against truthful and heuristic strategies. Unlike heuristics, LLMs leverage historical outcomes and prompt-based reasoning to adapt their bidding behavior dynamically. Results show that when the theoretical assumptions guaranteeing truthfulness hold, LLM bidders recover near-equilibrium outcomes consistent with VCG predictions. However, when these assumptions break -such as under static budget constraints- LLMs sustain longer participation and achieve higher utilities, revealing their ability to approximate adaptive equilibria beyond static mechanism design. This work provides the first systematic evaluation of LLM bidders in repeated spectrum auctions, offering new insights into how AI-driven agents can interact strategically and reshape market dynamics in future 6G networks.
\end{abstract}


\section{Introduction}
\subsection{Background}
Sixth-generation (6G) networks are expected to support massive connectivity and ultra-reliable low-latency services, with vehicular networks playing a key role in cooperative perception, autonomous driving, and real-time V2X communication~\cite{Walid_2019_MNET}.
Efficient spectrum allocation is central to these systems, where a base station (BS) distributes limited channels among multiple user equipments (UEs) competing based on private valuations and budget constraints. Designing intelligent bidding strategies is thus vital for maintaining fairness and efficiency in future wireless markets.

Auctions have long been effective for spectrum allocation~\cite{Yang_2013_COMST_AuctionTheory, Ismail_2021_Globecomm}, ensuring allocative efficiency where resources go to users who value them most. Among them, the Vickrey–Clarke–Groves (VCG) mechanism stands out for its incentive compatibility and efficiency in static, unconstrained environments. However, traditional formulations often assume one-shot interactions and truthful users, neglecting dynamics such as changing channel conditions, budget depletion, and evolving participation.
This naturally gives rise to repeated auction settings, where spectrum resources are re-allocated periodically, reshaping equilibrium behavior~\cite{ HanZhu_2011_TWC_RepeatedAuctions}.
Understanding how classical mechanisms perform in these settings, and evaluating how different bidding strategies such as truthful bidding, heuristic shading, or learning-based policies affect outcomes, is thus an open and practically relevant research problem, which is addressed in this paper.

\subsection{Related Works}
Emerging vehicular networks are characterized by their highly dynamic topologies, heterogeneous connectivity infrastructures, and fluctuating traffic demands. These conditions make centralized resource allocation schemes increasingly inadequate, as they struggle to cope with fast-changing link qualities, vehicular mobility, and real-time service constraints. 
Spectrum scarcity and co-channel interference among vehicles, roadside units (RSUs), and base stations further emphasize the need for distributed, market-driven access mechanisms~\cite{Yang_2013_COMST_AuctionTheory}.
For instance, a Vickry auction was designed in~\cite{Chang_TVT_2010} for spectrum reuse in vehicular networks.

With the emergence of Artificial General Intelligence (AGI) and the widespread deployment of \acp{LLM}, future wireless systems must anticipate environments where intelligent agents are no longer confined to the network side but also operate at the user end~\cite{Walid_2025_AGI}. 
A recent work by Microsoft~\cite{mao_2024_Alympics_LLM} analyzed the use of LLMs for a water allocation problem modeled as a first-price auction with budget constraints, but without valuation limits, and was restricted to the allocation of a single item per round. In their setting, each bidder was equipped with an LLM module that guided strategic bidding decisions based on budget availability, urgency requirements, and predefined personas. This provided valuable insights into the potential of LLMs as autonomous economic agents in competitive auction markets. However, that setting involved a single-item allocation and ignored valuation limits, unlike real-world spectrum markets where bidders are valuation-bounded and compete over multiple heterogeneous channels.

\subsection{Contribution}
Motivated by this, we envision reasoning-enabled vehicular agents -powered by lightweight \acp{LLM}- capable of autonomously participating in local spectrum auctions and making context-aware, economically rational decisions about bidding and association under budget constraints.
In our formulation, UEs are subject to both budget and valuation limits, and they may adopt different bidding strategies: (i) truthful bidding, (ii) heuristic bid shading, or (iii) LLM-based bidding. We focus on the widely studied VCG mechanism, combining theoretical benchmarks with simulation-based analysis to evaluate how LLM agents adapt their bidding across repeated rounds. This extension addresses two key challenges absent in prior works: enforcing valuation-bounded bidding behavior in repeated auctions and handling allocation across multiple spectrum channels.
Our work explores how LLMs, acting as reasoning agents, can navigate this richer game-theoretic landscape, approximating equilibrium strategies through adaptive decision-making rather than fixed analytical solutions.

Our findings uncover both expected consistencies with classical auction theory and important deviations that emerge under budget dynamics. In VCG auctions, when the standard assumptions for truthfulness hold -such as unlimited budgets-, LLM-based bidders achieve nearly identical utilities and win rates as truthful bidders, confirming that incentive compatibility remains robust even without explicitly enforcing it in the agents. However, when these assumptions are relaxed (i.e., under static budget constraints and repeated settings) LLM bidders adapt by pacing their spending, thereby sustaining participation over more rounds and achieving higher utilities than truthful or heuristic agents. Moreover, in scenarios where all UEs employ LLM-based bidding, the BS’s utility drops sharply compared to other strategies, indicating that structural modifications are needed for the BS to sustain acceptable profits in the presence of intelligent agents.
These results highlight three key contributions: (i) LLM agents can naturally recover classical equilibrium behavior in settings where it is theoretically guaranteed, and (ii) they can uncover adaptive, utility-enhancing strategies when constraints alter theoretical guarantees, and (iii) they push the boundaries of current auction mechanisms by exposing vulnerabilities and motivating the design of more robust, budget-aware, and strategic-aware allocation protocols for dynamic wireless networks. At a broader level, this demonstrates the potential of LLM-driven agents not only to validate existing allocation mechanisms, but also to stress-test their robustness, revealing opportunities for more resilient auction design in 6G networks and inspiring future research on adaptive AI mechanisms for dynamic spectrum sharing.

The rest of the paper is organized as follows. In Section~\ref{sec_system_model} we present our studied system and define the evaluation metrics. In Section~\ref{sec_pb_formula} we derive the different channel allocation procedures. We present our results in Section~\ref{sec_results} and conclude the paper in Section~\ref{sec_conclusion}.

\section{System Model}\label{sec_system_model}

Here, we formalize the repeated budget-constrained auction model, where a single BS allocates multiple channels to UEs with evolving budgets and valuations.

\subsection{System Description}
We consider a repeated spectrum auction framework for a vehicular 6G wireless network, where a single BS (e.g., RSU) acts as the auctioneer and a set of UEs act as bidders. The BS has $K$ finite number of orthogonal channels to allocate in each round, and the UEs compete to acquire one or more channels to satisfy their traffic demands. 
The allocation and payments are then determined by the chosen auction mechanism (e.g., \ac{VCG}). 
Let $\mathcal{S}$ denote the set of sub-channels available for allocation for the UEs. We assume that the BS transmits over sub-channel $k \in \mathcal{S}$ with a fixed power level denoted by $P$. 
Let $\mathcal{U}$ be the set of UEs located within the service area and $\Psi =\{\psi_1, \psi_2,\dots, \psi_i, \dots, \psi_{|\mathcal{U}|}\} $ denote the set of budgets for each UE. Each UE $i \in \mathcal{U}$ requests a downlink data rate $\overline{R}_i \in \Omega$, where $\Omega$ is a predefined discrete set representing the available service classes (in bits per second)\footnote{Note that, hereinafter, the terms service classes and \ac{QoS} levels are used interchangeably.}.
This setup captures the essential trade-offs in future vehicular 6G networks, where spectrum auctions will occur repeatedly, budgets are finite, and adaptive AI-driven bidding strategies may emerge. The following subsections formalize the auction process, bidder models, and performance metrics used in our proposed framework. Commonly used notations are provided in Table~\ref{table:00}.


\begin{table}[ht!]
\begin{center}
\caption{Table of Commonly Used Notations}
\begin{tabular}{ ||m{2.0cm}|m{5.5cm}|| }
 \hline
  Notation & Description \\ 
 \hline\hline
 $K$ & The number of available sub-channels  \\ 
 \hline
 $P$ & BS transmit power \\ 
 \hline
 $\mathcal{U}$ & The set of UEs \\ 
 \hline
 $\Psi$ & The set of budgets for each UE \\ 
 \hline
  $\overline{R}_i$ & Requested downlink rate by UE $i$ \\ 
\hline
 $\Omega$ & The set of services classes \\ 
 \hline
  $R_i$ & The achievable downlink rate for UE $i$ \\ 
 \hline
 $\hat{R}_i$ & The total achievable downlink rate for UE $i$\\ 
 \hline
 $\gamma_i$ & The SINR or UE $i$ \\ 
 \hline
 $\eta$ & The resource demand ratio \\ 
 \hline
 $\kappa_i$ & The submitted bid by UE $i$ \\ 
 \hline
 $r$ & The reservation price for 1 sub-channel \\ 
 \hline
 $u_i$ & The utility of UE $i$ \\ 
 \hline
 $\hat{u}_i$ & The utility of the BS for serving UE $i$ \\ 
 \hline
 $v_i$ & The valuation of UE $i$ for all the allocated sub-channels \\ 
 \hline
 $\mathcal{A}^*$ & The set of UEs that maximizes total welfare\\
 \hline
\end{tabular}
\label{table:00}
\end{center}
\end{table}

\subsection{Evaluation Metrics: Data Rate and QoS}

The achievable data rate $R_{i}$ for UE $i$ on sub-channel $k$ is modeled using the Shannon capacity formula as~\cite{ThantZin_2017_TMC}:

\begin{equation}\label{eq_data_rate_k}
\small
R_{i} = W \cdot \log_2(1 + \gamma_i),
\end{equation}

\noindent where $W$ is the bandwidth of a single sub-channel and $\gamma_i$ is the instantaneous \ac{SINR} at UE $i$, when served by the BS on sub-channel $k$.
The total data rate $\hat{R}_{i}$ achieved by UE $i$ across all its allocated $N_{i}$ sub-channels from the BS is given by:

\begin{equation}\label{eq_data_rate_total}\small
\hat{R}_{i} =  \sum_{k \in \mathcal{S}} y_{i}^{(k)} \cdot R_{i}.
\end{equation}

\noindent where $y_{i}^{(k)} \in \{0,1\}$ indicates whether sub-channel $k$ is allocated to UE $i$.
Additionally, for assignment feasibility, a minimum QoS should be guaranteed by the BS, which is translated into the following constraint:
\begin{equation}\label{eq_QoS_constraint}\small
    \hat{R}_{i} \geq \overline{R}_i, \quad \forall i \in \mathcal{U}.
\end{equation}

\textcolor{black}{Note here that constraint~\eqref{eq_QoS_constraint} is a pre-auction, UE-side feasibility check where each UE independently verifies whether its service requirement can be met by a given BS-sub-channel configuration and, based on this verification, determines the minimum number of sub-channels $N_i$ required to satisfy its target data rate.}
We also define the resource–demand ratio as: $\eta~=~\frac{K}{D}$, where $D=\sum_{i \in |\mathcal{U}|}N_i$ is the total demand from all UEs. Three operating regimes can be captured here: scarcity ($\eta < 1$), where demand exceeds supply and competition intensifies; balanced ($\eta \approx 1$), where supply and demand are roughly matched; and abundant ($\eta > 1$), where resources are plentiful and competition diminishes.

\subsection{Utility Functions}
\subsubsection{BS Utility}

Since electricity consumption constitutes a significant portion of a base station’s operational cost, we model the downlink operational cost primarily in terms of energy expenditure. Specifically, we define the operational cost incurred by the BS when transmitting on a sub-channel as:

\begin{equation}\label{eq_reservation_price}\small
    r = \mu P,
\end{equation}

\noindent where $\mu$ is the unit price of transmission power at the BS.
The operational cost defined in~\eqref{eq_reservation_price} is considered as the reservation price (i.e., minimum price) set by the BS for each sub-channel. 
The utility of the BS from serving UE $i$ is then defined as the difference between the payment received from UE $i$ and the operational cost, which we denote formally as:

\begin{equation}\label{eq_utility_BS}\small
   \hat{u}_i = N_{i} (\pi_{i} - r),
\end{equation}

\noindent where $N_i$ is the number of requested sub-channels and $\pi_{i}$ is the payment of UE $i$ for each allocated sub-channel.
By maximizing its overall utility, the BS is able to achieve higher value from allocating its sub-channels.

\subsubsection{UE Utility}

The utility of UE $i$ is defined as the difference between its valuation of all the allocated sub-channels and the payment given to the BS. Formally, this is expressed by the following quasilinear preference
function:

{\small \begin{equation}\label{eq_utility_UE}\small
u_{i} = \begin{cases} 
N_{i} (v_{i} - \pi_{i}), & \text {if UE $i$ wins},
\\ 0, & \text {otherwise}.
\end{cases}
\end{equation}}

\noindent where $v_{i}$ is the per-channel valuation of UE $i$. The valuation is formally defined as:
\begin{equation}\label{eq_v_i}\small
    v_{i} = \alpha_i R_{i},
\end{equation}

\noindent where $\alpha_i$ is a UE-specific scaling parameter converting rate to utility (e.g., importance of throughput) and $R_{}$ is the achievable data rate defined in~\eqref{eq_data_rate_k}. 
By maximizing its overall utility, the UE is able to allocate sub-channels at a lower cost. These contradicting objectives between the BS and UEs naturally create a competitive environment where strategic interactions shape outcomes. Next, we formulate the problem based on auction mechanism, which provide a principled framework to balance these incentives, ensuring efficient allocation while preserving fairness and revenue considerations.

\section{Problem Formulation: Auction-based Channel Allocation}\label{sec_pb_formula}

In the following, we first describe two techniques that the UEs employ to select the optimal bid value to maximize their utility. We then define the BS-side channel allocation procedure, highlighting the use of the \ac{VCG} auction model to determine the final allocations and prices.

\subsection{Bid Value Selection Procedure (UE-Side Decision)}
A straightforward strategy in spectrum auctions is for each UE to directly submit its true valuation as the bid. This approach is natural in VCG mechanisms, where truthfulness is a dominant strategy in single-shot, budget-free settings. However, in 6G repeated auctions with budget and urgency constraints (e.g., 6G-connected autonomous vehicle with low-latency links for safety-critical updates), strictly truthful bidding may no longer maximize long-term utility, leaving no resources for later episodes when communication is still required. This motivates the use of heuristic bidding rules, such as shaded bidding, which aim to preserve budget while maintaining competitiveness. Yet, designing effective heuristics is challenging, as fixed shading parameters can be too rigid to adapt to dynamic environments. In this context, LLM-based bidding represents a promising alternative: by reasoning over valuation, history, and budget, LLM agents are expected to generate adaptive bidding strategies that bridge the gap between rigid heuristics and theoretical benchmarks. Here we describe both the heuristic and LLM-based approaches.

\subsubsection{Heuristic-based Bidding}

The shaded bidding heuristic interpolates between the UE’s true valuation and the most recently observed clearing price. The idea is to remain conservative when the UE has sufficient remaining opportunities to win, but to bid more aggressively as the number of unsuccessful rounds without channel allocation increases. Formally, let $\pi_{\text{clr}}$ denote the most recent clearing price, $f^{(t)}_i$ be the number of consecutive failures experienced by the user at time slot $t$, and $F_{\max}$ be the maximum tolerated number of failures before the user is temporary blocked from future auctions. The bidding rule to calculate the bidding value $\kappa^{(t)}_i$ at time step $t$ is then defined as:
\begin{equation}\small
    \kappa^{(t)}_i = \beta^{(t)}_i v_i + (1-\beta^{(t)}_i) \pi_{\text{clr}} , 
\end{equation}

\noindent where $\beta^{(t)}_i = \frac{\log(f^{(t)}_i+1)}{\log(F_{\max})}$ and $\beta^{(t)}_i \in [0,1]$.
By construction, $\beta^{(t)}_i$ increases with $f^{(t)}_i$, ensuring that as the user approaches $F_{\max}$ (i.e., the risk of being left unallocated), the bid converges to the true valuation $v_i$. This adaptive shading mechanism balances budget preservation with urgency, creating a smooth transition from conservative (i.e., $\pi_{\text{clr}}$) to aggressive (i.e., $v_i$) bidding behavior at different times. 
\textcolor{black}{
Note here that as the clearing price $\pi_{\text{clr}}$ might exceed the UE's valuation $v_i$ (e.g., due to the presence of other high valuation UEs), the final bid value $\kappa_i$ should be capped as $\kappa^{(t)}_i = min\{v_i, \kappa^{(t)}_i\}$.
}

\subsubsection{LLM-based Bidding}

To enable more sophisticated reasoning, we propose an approach in which the UEs are equipped with an LLM module to assist in bid estimation, based on their local observations and constraints. The LLM-based UE formulates a structured prompt containing the required number of sub-channels $N_{i}$, its own budget $\psi_i$, and the spectrum valuation $v_{i}$. This prompt is submitted to the LLM module of the UE, which returns a tuple ($\kappa_i, \mathcal{E}_i$) indicating the suggested bid value $\kappa_i$ and a textual explanation $\mathcal{E}_i$ of the reasoning. 
Over repeated auction rounds, the UE refines its prompts by including outcome feedback (e.g., previous clearing prices, previous own bids and auction outcomes), enabling adaptive learning. 
Figure~\ref{fig:llm_prompt} illustrates a template of an LLM prompt and the expected response.

\begin{figure}[t]
\begin{tcolorbox}[colback=black!10, colframe=black!50, width=1\linewidth, boxrule=0.5pt]
\centering
\fbox{
\begin{minipage}{0.95\linewidth}
\tiny
\textbf{LLM Prompt Template:}

\texttt{
Given the following network and economic context: \\
- Your true valuation for the BS spectrum. \\
- Your budget: $\beta_i$ \\
- Number of sub-channels required : $N_{i}$ \\
- Previous clearing prices.\\
- Previous own bids and auction outcomes.\\
Please analyze and provide: \\
1. Recommended bid value for the spectrum. \\
2. A brief explanation of your reasoning.
}

\textbf{Expected Response Format:}

\texttt{Bid value: [value] \\
Explanation: "[Short textual reasoning]"
}
\end{minipage}
}
\end{tcolorbox}\caption{Template of the LLM prompt and expected response structure used for strategic bidding.}\label{fig:llm_prompt}
\end{figure}

\subsection{Channel Allocation Procedure (BS-Side decision)}
Once the bid value is derived, UE $i$ transmits a resource request to the BS containing the number of sub-channels required and the offered price.
The allocation of sub-channels at the BS is managed through an auction mechanism that is executed periodically (i.e., every allocation window $T_c$).
In VCG auctions, each winning UE pays a price equivalent to the externality it imposes on others, encouraging truthful bidding. \textcolor{black}{Formally, the BS selects an allocation $\mathcal{A}^*$ that maximizes the system's total welfare  as follows:
\begin{equation}\small
    \mathcal{A}^* = \arg\max_{\mathcal{A} \subseteq \mathcal{F}} \sum_{i \in \mathcal{A}} N_{i}(\kappa_i - r) \quad \text{s.t.} \quad \sum_{i \in \mathcal{A}} N_{i} \leq K,
\end{equation}
\noindent where $\mathcal{F} \subseteq \mathcal{U}$ is the set of UEs that submitted their requests to the BS.
The payment for UE $i$ is then derived as
\begin{equation}\label{eq_p_i}\small
    p_i = \sum_{j \in \mathcal{A}^{-i}} N_{j}(\kappa_j - r) - \sum_{j \in \mathcal{A}^* \setminus \{i\}} N_{j}(\kappa_j - r),
\end{equation}
\noindent where the first term in~\eqref{eq_p_i} is the optimal social welfare in the  absence of UE $i$ and the second term is the social welfare of all other UEs when UE $i$ participates in the auction.}
In single-shot VCG auction with no budget constraint, bidding the true valuation is a dominant strategy. However, the incorporation of budget constraints into our repeated auction system makes truthful biding no longer the optimal strategy, as we show in the next section.

\section{Experimental Analysis}\label{sec_results}


\textcolor{black}{
We present in this section simulation results comparing truthful, heuristic-based, and LLM-based bidding strategies in repeated spectrum allocation auctions. Performance is evaluated using two key metrics: (i) winning frequency and (ii) accumulated utility, jointly reflecting allocation success and economic efficiency. Unless otherwise stated, the setup involves 16 UEs competing for 6 sub-channels ($\eta < 1$) over 20 auction episodes. 
Similar to the work in~\cite{Hu_2024_AAMAS}, to evaluate the performance of the LLM bidding, we consider an opponent that is consistently using the truthful bidding.
UE 10 adopts truthful bidding and UE 13 employs LLM-based bidding while the remaining UEs follow heuristic bidding.
The sub-channel bandwidth is set to $W = 180$\,kHz (consistent
with a standard 5G/NR numerology that carries over to 6G
sub-6\,GHz deployments~\cite{Walid_2019_MNET}) and
$\gamma_i \sim \mathcal{U}[5, 20]$\,dB captures the range of
instantaneous SINR values experienced by vehicular UEs in a
mixed urban/suburban scenario.
Each UE's scaling factor is drawn as $\alpha_i \sim
\mathcal{U}[0.8,\,1.2]$, yielding per-channel valuations
$v_i = \alpha_i R_i \in [1.0,\,3.5]$ monetary units,
consistent with the valuation intervals adopted in~\cite{HanZhu_2011_TWC_RepeatedAuctions}.
The reservation price is set to $r = \mu P$, where $P = 200$\,mW
is the BS transmit power and $\mu = 6$\,monetary units/W
represents a typical small-cell energy price, giving $r \approx 1.2$.
For the LLM-based strategy, we employ the \texttt{gpt-5-mini} model.
}

\textcolor{black}{
Results are organized around two distinct budget dynamics, each of which provides different insights into the role of strategy and learning in repeated auctions.
In the \emph{budget-refill} scenario, each UE's budget is refreshed
every episode by a draw from $\mathcal{U}[r-\epsilon,\, r+\epsilon]$
with small $\epsilon$, modelling episodic micro-payment billing
(e.g., per-coherence-time spectrum credits in network slicing),
where each credit suffices for roughly one channel reservation.
In the \emph{static-budget} scenario, each UE is assigned a
fixed budget $\psi_i = 15$, which represents approximately
$15/r \approx 12.5$ episode-equivalents of the reservation price,
capturing a prepaid spectrum-credit model in which UEs must
plan their spending across the full $T=20$-round horizon.
}


\begin{figure}[!h] 
     \centering
     \begin{subfigure}[b]{0.24\textwidth}
         \centering         \includegraphics[width=\textwidth,height=3.0cm]{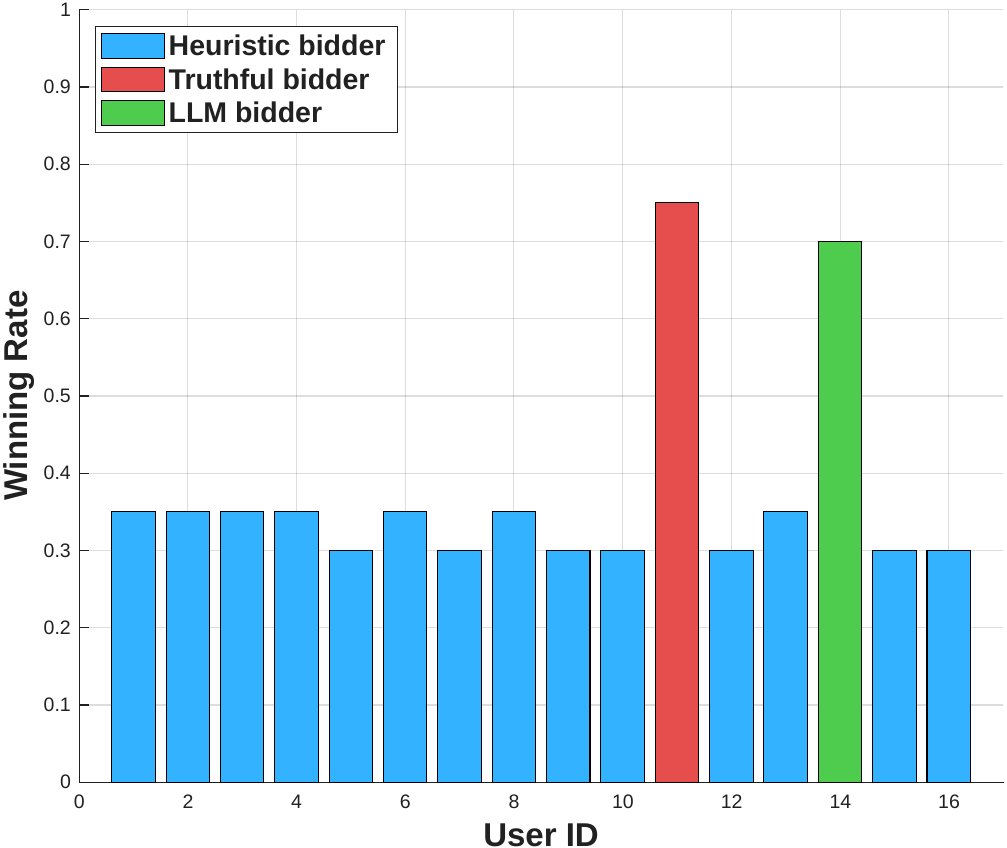}
         \caption{Winning frequency.}
         \label{fig:vcg_win_rate}
     \end{subfigure}
     \hfill
     \begin{subfigure}[b]{0.24\textwidth}
         \centering         \includegraphics[width=\textwidth,height=3.0cm]{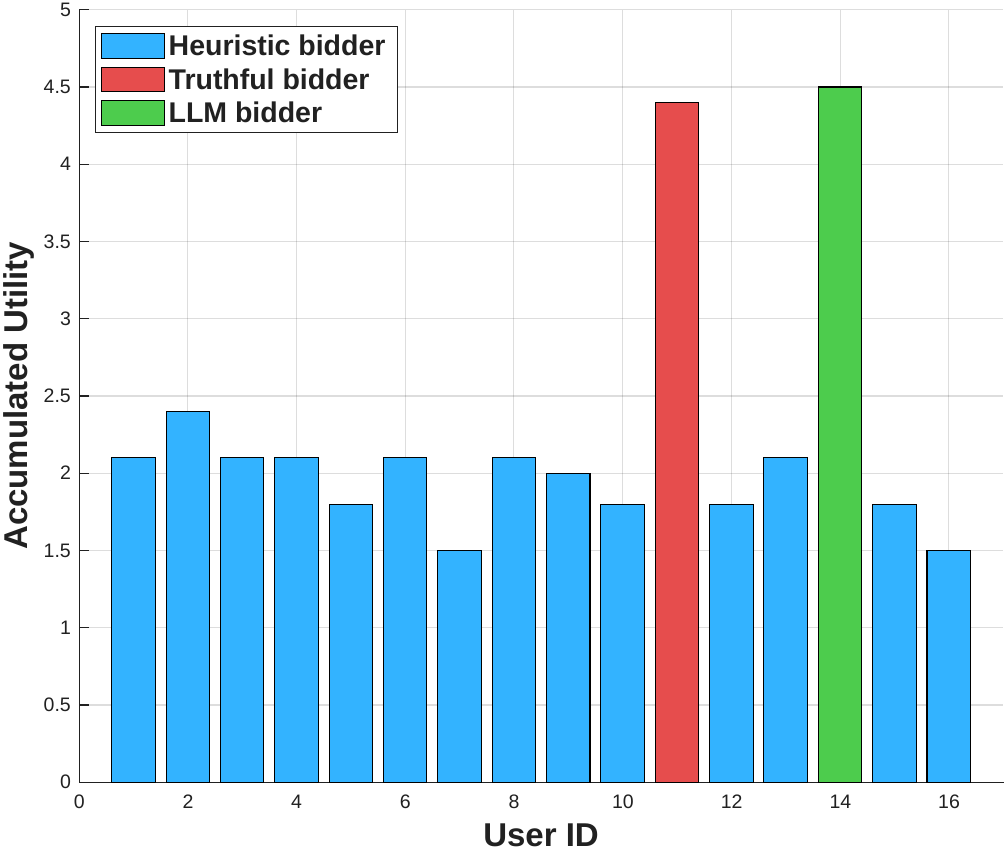}
         \caption{UEs utility.}
         \label{fig:vcg_acc_utility}
     \end{subfigure}
        \caption{Winning frequency and utilities with budget refill.
        }
        \label{fig:budget_1_win_and_utility}
\end{figure}

\begin{figure}[!h] 
     \centering
     \begin{subfigure}[b]{0.24\textwidth}
         \centering         \includegraphics[width=\textwidth,height=3.0cm]{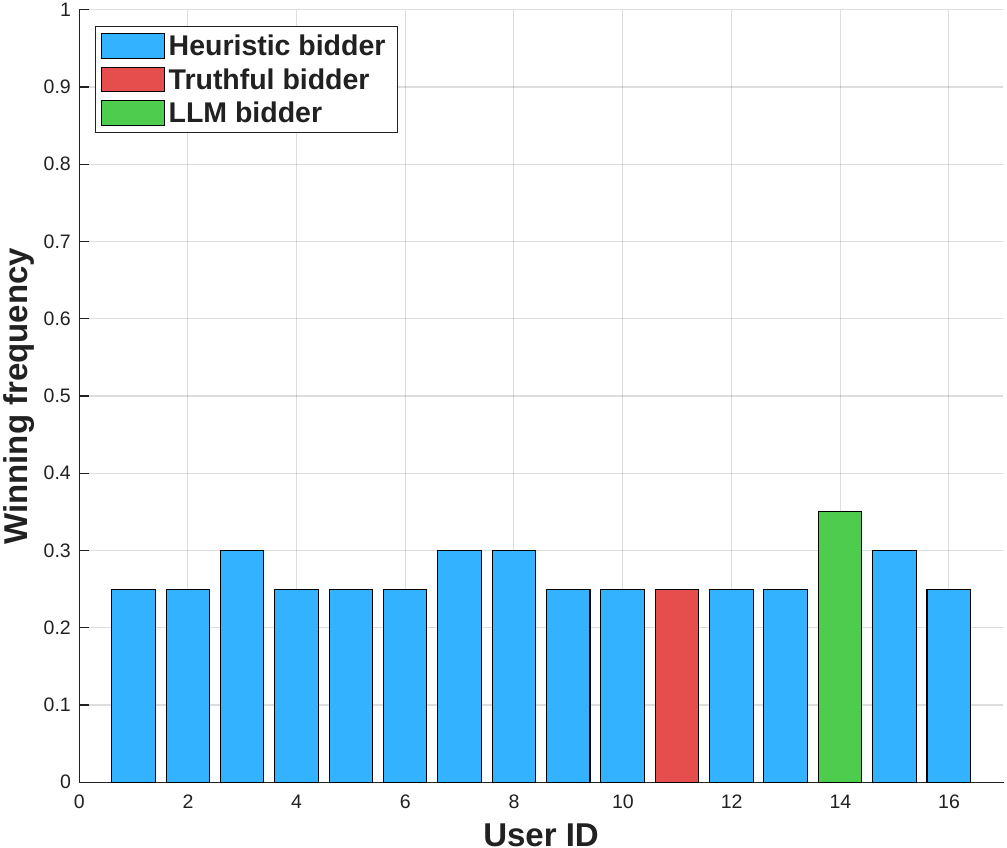}
         \caption{Winning frequency.}
         \label{fig:budget_2_vcg_win_rate}
     \end{subfigure}
     \hfill
     \begin{subfigure}[b]{0.24\textwidth}
         \centering         \includegraphics[width=\textwidth,height=3.0cm]{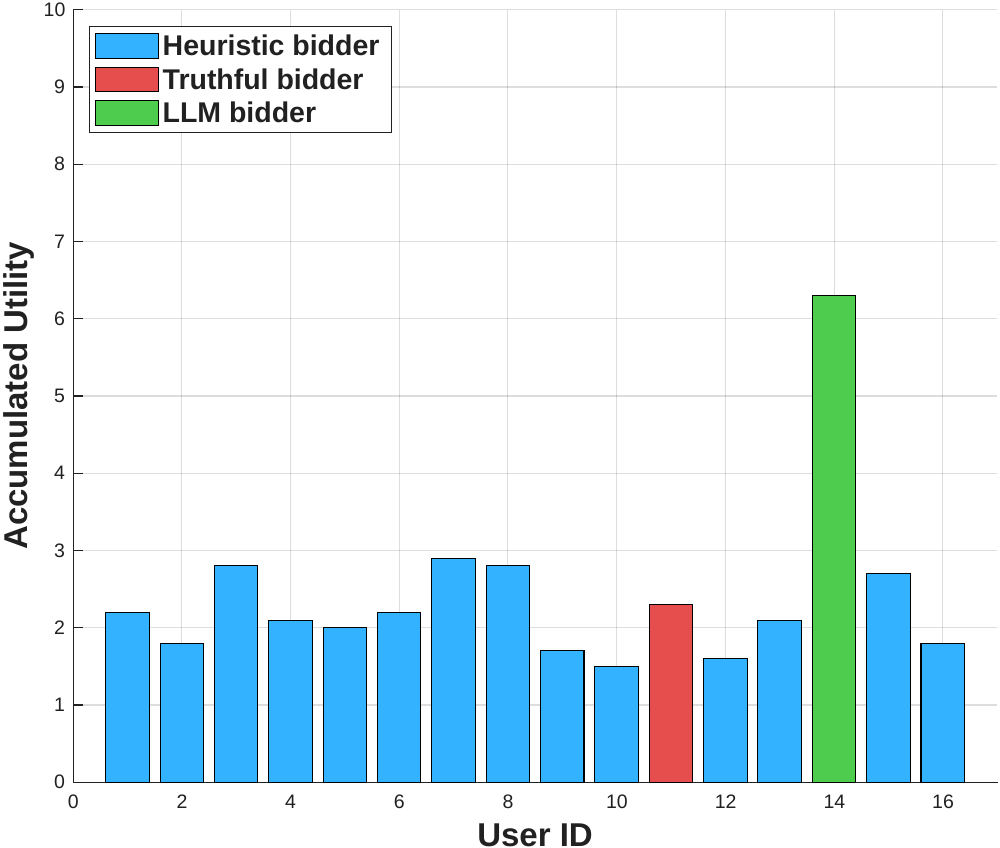}
         \caption{UEs utility.}
         \label{fig:budget_2_vcg_acc_utility}
     \end{subfigure}
        \caption{Winning frequency and utilities with static budget.
        }
        \label{fig:budget_2_win_and_utility}
\end{figure}

\subsection{Constantly Refilled Budgets (Quasi-Unconstrained Setting)}
Figures~\ref{fig:vcg_win_rate} and ~\ref{fig:vcg_acc_utility} present the winning frequency and accumulated utility of each UE over the 20 auction episodes, respectively, for the \emph{budget-refill} scenario. 
The results reveal that the LLM-based bidder closely mirrors the truthful bidder, achieving nearly identical win rates and utilities.
The budget management approach here effectively models a scenario where users always have sufficient resources to participate in every round, and hence the repeated auction reduces to a sequence of independent single-shot games~\cite{wang2023learning}. In this regime, the theoretical properties of the underlying auction mechanisms dominate: in VCG auctions, truth-telling is strictly optimal regardless of repetition, and as expected, the LLM bidder is unable to outperform a straightforward truthful strategy.

\subsection{Fixed Budgets (Budget-Constrained Setting)}
The second set of simulations examines a more realistic case with \emph{budget-static} scenario, making the budget constraint more binding~\cite{wang2023learning}. Unlike the previous setting, repetition here introduces a strong intertemporal dimension: bidders must pace their expenditure to sustain participation across multiple episodes. 
Here we adjust the LLM's prompt by requiring the agent to \textit{``maximize cumulative utility while never exhausting the budget before the last episode"}.
Figures~\ref{fig:budget_2_win_and_utility} present the results of this experiments.
The LLM agent is now able to clearly outperform the truthful bidder in the VCG auction. 

This budget management approach fundamentally changes the equilibrium dynamics, since aggressive early bidding leads to premature exit from the market, while conservative pacing creates opportunities to benefit from reduced competition in later episodes.
Specifically, the final winning episodes occur, on average, around the 4th round for truthful bidding, the 12th for shaded bidding, and the 17th for LLM-based bidding.
These results show that the LLM bidder achieves superior performance compared to both truthful and heuristic strategies. 
As UEs interact repeatedly, the LLM-based bidder implicitly converges toward a form of dynamic equilibrium, balancing immediate utility against future opportunities.
This is achieved by striking a balance between competitiveness in early rounds and budget preservation for later episodes, thereby maintaining participation until later stages of the game. Importantly, this translates into higher cumulative utility: as weaker bidders deplete their budgets and exit, the market becomes less competitive, driving down clearing prices and allowing the LLM bidder to capture more value at a lower cost.

\subsection{Impact of Resource–Demand Imbalance on Strategic Bidding Performance}
\begin{figure}[htbp]
    \centering    \includegraphics[width=0.7\columnwidth,height=3.1cm]{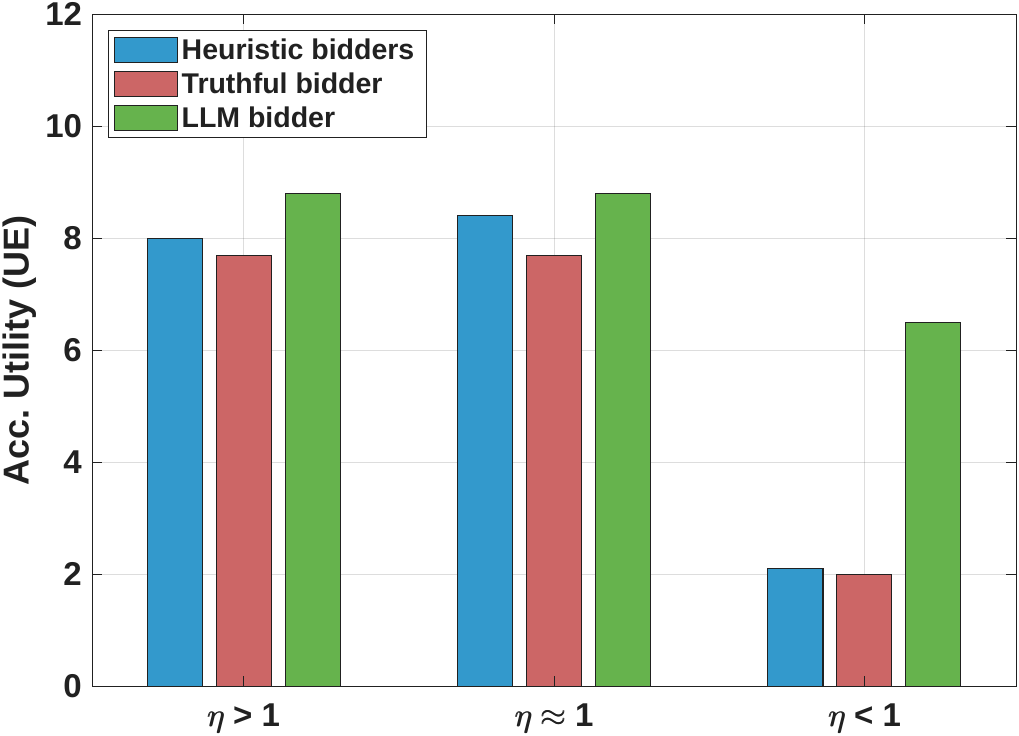}
    \caption{Accumulated UE utilities for different resource-demand ratio.}
    \label{fig:vcg_utility_scarcity}
\end{figure}

As shown in Figure~\ref{fig:vcg_utility_scarcity}, when the total available spectrum exceeds the user demand, all bidding strategies yield identical utilities, confirming that the VCG mechanism collapses to the reservation price regime -a well-established result in auction theory practice~\cite{krishna2009auction}. In this low-competition setting, every UE receives the required sub-channels, and payments equal the reserve price since no bidder imposes externalities on others. However, as the gap narrows and demand begins to exceed supply (i.e., $\eta < 1$), the auction transitions into a competitive regime. In this region, the benefits of strategic reasoning become evident: LLM-guided UEs adaptively conserve their budgets and selectively participate in high-value rounds, achieving superior long-term utility and access frequency compared to both truthful and heuristic bidders.
Note here that the overall utility drops as competition intensifies because increased demand raises clearing prices and limits each user's chance of winning, thereby reducing the individual surplus obtained per round.

\subsection{Impact on BS utility}
\begin{figure}[!h] 
     \centering
     \begin{subfigure}[b]{0.24\textwidth}
         \centering         \includegraphics[width=0.8\textwidth,height=2.6cm]{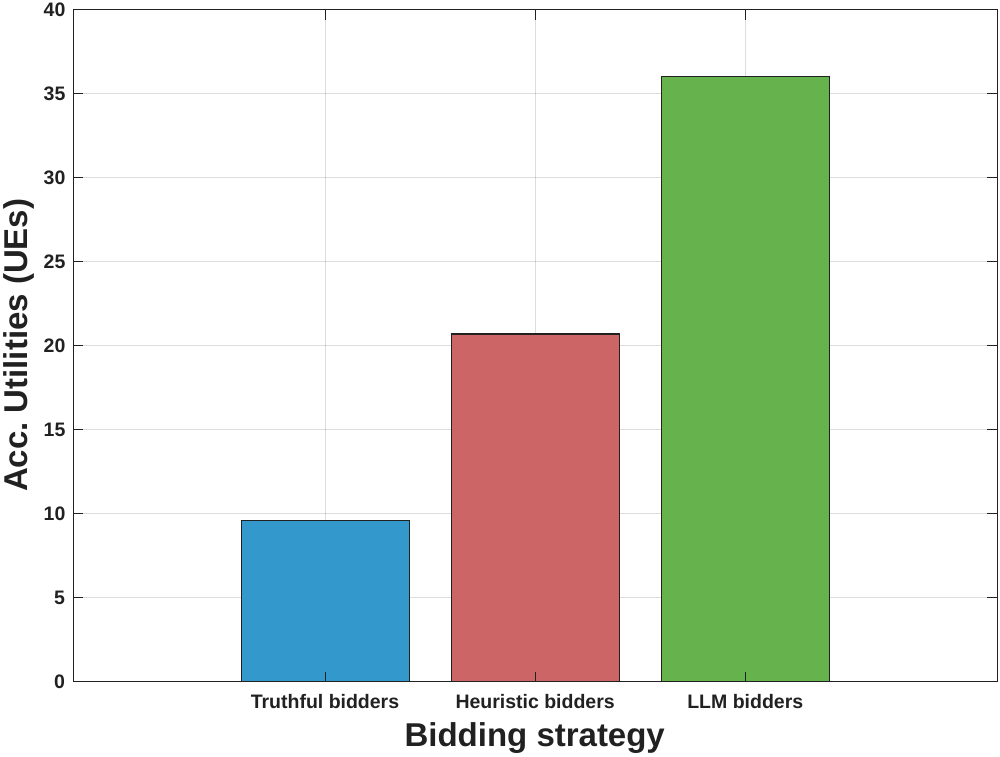}
         \caption{}
         \label{fig:acc_utilities_UEs}
     \end{subfigure}
     \hfill
     \begin{subfigure}[b]{0.24\textwidth}
         \centering         \includegraphics[width=0.8\textwidth,height=2.6cm]{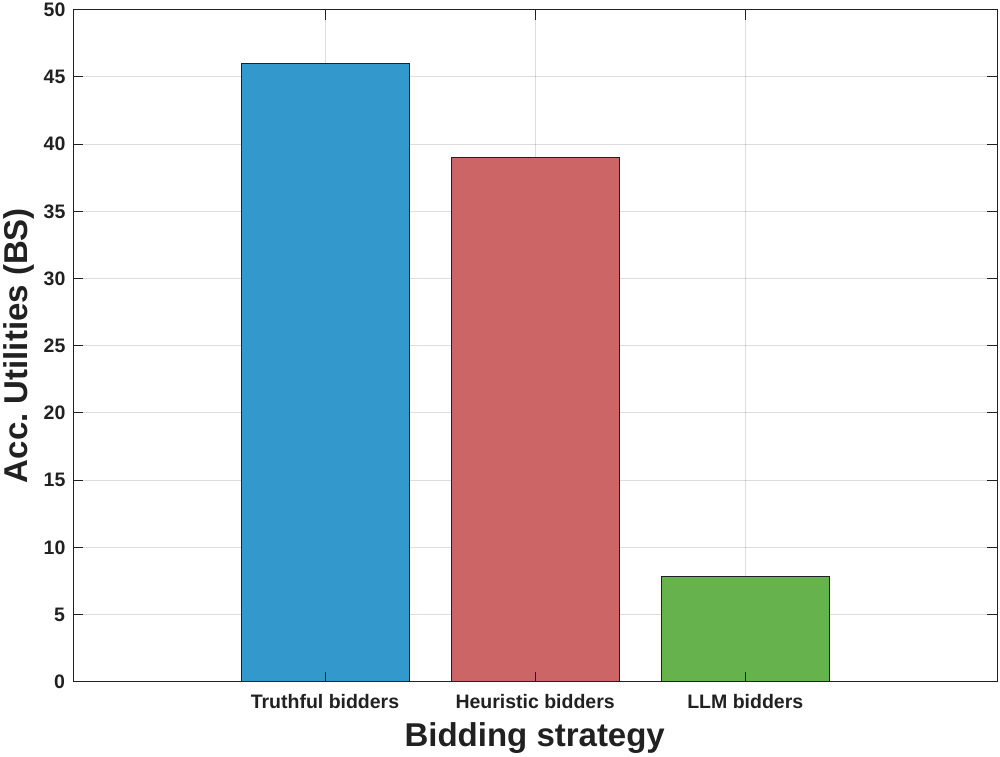}
         \caption{}
         \label{fig:acc_utilities_BS}
     \end{subfigure}
        \caption{Accumulated utilities for (a) UEs and (b) BS with static budget.
        }
        \label{fig:acc_utilities_BS_and_UEs}
\end{figure}

To illustrate the impact of LLM-based bidding, we consider three extreme cases in a static budget setting: a) all truthful, b) all heuristic, and c) all LLM-based bidders. As shown in Figure~\ref{fig:acc_utilities_BS_and_UEs}, results reveal a clear inverse relationship: as the sophistication of the bidding strategy increases, UEs capture more surplus, leaving the BS with significantly lower utility. In particular, the use of LLMs enables UEs to reason strategically over the budget and competition, maximizing their gains while sharply reducing the BS’s revenue compared to the all-greedy scenario. This outcome underscores the need for future work to explore adaptive auction mechanisms or counter-strategies that allow the BS to maintain revenue in the presence of intelligent, learning-driven bidders.

\subsection{Discussions}
\textcolor{black}{
In the area of \ac{RTB} for online advertising, the problem of bidding repeatedly under budget constraint has been heavily investigated~\cite{Ou2023_survey_bid, Aggarwal2024_survey_autobidding}.
Nevertheless, when viewed from the perspective of wireless communications, the problem we consider exhibits several fundamental differences. Specifically, within the wireless communications community, the VCG mechanism has been the predominant auction framework adopted for spectrum and resource allocation due to its strong theoretical properties. In RTB literature, generalized second-price (GSP) and generalized first price (GFP) auctions are the more commonly used schemes. In this context, our study align with the direction of the community to showcase that VCG should not be always used blindly as it has some practical limitations. While VCG is theoretically appealing, it exhibits important limitations in environments characterized by budget constraints, repeated interactions, and strategic user behavior. Demonstrating how these limitations manifest, and exploring alternative bidding behaviors under more realistic assumptions, constitutes a central contribution of this work to the wireless communications literature.
}

\textcolor{black}{
Having said that, the heuristic bidding strategy adopted in this work can be viewed as a specific instantiation of the linear bidding frameworks widely studied in the RTB literature (e.g., as in~\cite{Ou2023_survey_bid, Yang2019_AiAds}). Under linear bidding, the submitted bid is expressed as a scaled (i.e., shaded) version of the bidder's true valuation $v_i$, where the scaling factor captures budget considerations, competition intensity, and pacing objectives. This factor may be determined through rule-based mechanisms~\cite{Yang2019_AiAds} or learned via data-driven approaches such as deep reinforcement learning~\cite{Lu2019_ileanrbidding}. While linear bidding does not generally yield a provably optimal solution from a theoretical standpoint, it has emerged as the dominant practical bidding paradigm in large-scale real-world systems due to its simplicity, robustness, and empirical effectiveness under dynamic and budget-constrained environments.
In single-shot, budget-unconstrained settings, the optimality of such linear scaling can in fact be established rigorously: using duality technique, \cite{Aggarwal_2019, he2021unified} show that a bid of the form $\kappa_i = w_0 v_i$ is derived from the optimal dual variable of the budget constraint, achieves near-optimal value if and only if the per-round auction is truthful.
}

\textcolor{black}{
However, this guarantee breaks down in our repeated, budget-constrained setting, where the depletion of the static budget across rounds introduces intertemporal incentives that violate per-round IC. Deriving an analogous offline near-optimal linear benchmark under repeated truthful auctions with depleting budgets -extending the duality framework of~\cite{Aggarwal_2019, he2021unified} to the 6G spectrum context- remains an open and theoretically non-trivial problem, which we identify as a direction for future work.
Finally, from a game theoretic perspective, it has be concluded in~\cite{Ou2023_survey_bid} that despite the existence of equilibria under mild conditions -for an optimal bidding strategy-, there is no general guarantee that adaptive bidding agents will converge to these equilibria in dynamic and non-stationary auction environments, suggesting that the heuristic methods are the more practical solutions.
}

\section{Conclusion}\label{sec_conclusion}

This paper investigated LLMs as strategic bidding agents in repeated 6G spectrum auctions under budget constraints. From a game-theoretic perspective, LLM bidders act as rational players capable of adaptive reasoning beyond dominant-strategy truthfulness. When VCG assumptions hold, LLMs recover near-equilibrium outcomes equivalent to truthful bidding, confirming mechanism robustness. Under static budgets, however, they pace expenditures, sustain participation, and achieve higher utility, revealing adaptive equilibria where incentive compatibility weakens. These findings suggest that LLMs can validate classical mechanisms while uncovering new strategic behaviors in constrained environments. \textcolor{black}{Future work will extend this analysis to other auction formats and compare LLM bidders with reinforcement learning-based strategies to guide the design of intelligent spectrum markets in 6G networks. Additionally, future mechanism design are expected to balance multiple objectives, including BS revenue and system-wide social welfare, while remaining robust to strategic and intelligent user behavior.}

\bibliographystyle{IEEEtran}
\bibliography{ref}

\end{document}